# Naturally self-assembled nickel nanolattice

Jaiveer Singh,[a] Netram Kaurav,[b] Niranjan Prasad Lalla[c] and Gunadhor Singh Okram[*d]

This is the first report on the critical nature of nanolattice formability of different particle size (~4-10nm) of monodispersed nickel nanoparticles. They exhibit strikingly hexagonal close-packed (hcp) nanolattices without extra forces, whenever trioctylphosphine is (one of) the surfactant(s). This clearly establishes the unique role of nanolattice formability of trioctylphosphine. The c/a ratios are interestingly identical to those of atomic lattices. An attempt has also been made to explain them based on the balanced attractive and repulsive forces of the surfactant-generated cation-anion pairs on the surface of the nanoparticles. The present findings therefore will provide a far-reaching vista to fabrication of varieties of natural nanolattices and their understanding on applications in a new paradigm.

## Introduction

The properties of nanoparticle lattices (nanolattices) are distinct from those of individual nanoparticle or bulk counterparts. They are highly attractive for future advanced applications[1-5] but have so far been enabled artificially using extra forces.[6-11] For example, tributylphosphine has been used for nickel nanolattice[6] formation similar to those of gold[12] and iron oxide.[13] Assembly of nanoparticles of two different materials into a binary nanolattice of varieties of materials (to enable opposite electrical charges on nanoparticles to impart a specific affinity of one type of particle)[2,11] or a nanolattice of oppositely charged nanoparticles[14] has been studied. However, such impressions of the compulsory use of an external agent seems to be due to the poor knowledge of surfactants in general (or specifically trioctylphosphine) since such so-called non-ionic surfactant is *usually ionic* due to the possible formation of cation-anion pairs from dissociated surfactant molecules or impurities.[15,16] This will then favor mimicking the naturally balanced electrostatic cohesive and repulsive energies of electrons and nuclei found in atomic lattices[17,18] without external forces and hence natural formability of nanolattice. This possibility, if proven, strongly suggests the paramount versatility of trioctylphosphine as one of the surfactants in the preparation of nanoparticles as it also most probably enables monondispersity.[6,19-25] This hypothesis also supports the probable unproven signature of CdSe nanolattice formation due to the presence of trioctylphosphine as one of the surfactants.[24] To critically test for natural nanolattice formability, we chose the case for nickel nanoparticles prepared from nickel acetylacetonate by fixing the content of one of the favorite surfactants or stabilizers among researchers (viz., oleylamine[6,19], trioctylphosphine[6,19-25] and triphenylphosphine[6]) while varying one of them or using only a single surfactant independently. Thus, we demonstrate (i) the formation of strikingly natural hcp nanolattices of nickel when no extra forces are used with the nanoparticles prepared as usual, whenever trioctylphosphine is (one of) the surfactant(s), (ii) the unqiue role of nanolattice formabilty of trioctylphosphine and (iii) the naolattice parameters, calculated analytically, to have c/a ratios identical to those of atomic lattices. These have been established concretely using, among others, small angle X-ray scattering (SAXS), transmission electron microscopy (TEM) and zeta potential techniques.

## Experimental

### Synthesis of monodispersed nickel nanoparticles

Thermal decomposition method as in references[6,19] was used to synthesize the nanoparticles. Typically, 1 ml (i.e. 2.24mM) of preheated (215 °C) trioctylphosphine (90% Aldrich) was added in the already degassed (at 100 °C for 30 min) solution of 1.02 g Ni(acac)$_2$ (95% Aldrich) and 8 ml oleylamine (70% Aldrich). The resulting solution was further heated at 220 °C for 2h under argon atmosphere. This gave rise to black precipitate due to formation of nickel nanoparticles. Solution was then cooled to 27 °C, and centrifuged by adding ethanol (99.9% Jiangsu Huaxi) to extract and wash the nanoparticles. Washing was done four times. Similar procedures were followed for 3ml, 5ml, 8ml and 10ml of trioctylphosphine at fixed (8ml) oleylamine; trioctylephosphine in X ml will denote the samples here. In addition to these samples, several other samples were prepared for (a) varying oleylamine with fixed trioctylphosphine, (b) varying triphenylphosphine (99% Aldrich) with fixed oleylamine and (c) separately for each of these surfactants. The particles were dried at 60 °C and used directly for characterizations.

### Synchrotron SAXS and XRD measurements

Synchrotron radiation (1.089Å) X-ray diffraction (XRD) data was collected at BL-18B (Indian beamline), Photon Factory, Tsukuba, Japan with a beam current of 401 mA in the angle ranges 0.2-2° and 9-30° for angular step of 0.025° with a point detector (Cyberstar) on powdered samples and glass drop-casted thin films. The thin films were made after thorough sonication of the nickel nanoparticles dispersed in ethanol. The incident X-ray angle for small angle X-ray scattering (SAXS) measurements was 0.15-0.25°.

### Laboratory X-ray Diffraction

The Bruker D8 Advance X-ray diffractometer with Cu K$\alpha$ radiation (0.154 nm) in the angle range 20-90° was used for laboratory method of XRD measurements of the samples in powder form; the X-rays were detected using a fast counting detector based on silicon strip technology (Bruker LynxEye detector).

### Laboratory high resolution SAXS

High resolution laboratory SAXS measurements on glass drop-casted thin films were done with Cu K$\alpha$ radiation in the angle range 0.2-10° with a step size of 0.02°; the incident X-ray angle was normally fixed at 0.5° unless it is specified.

### Transmission electron microscopy (TEM)

Nanoparticle images and selected area electron diffraction (SAED) were recorded using transmission electron microscopy (TECHNAI-20-G$^2$) by drop-casting the well-sonicated solution of a few milligrams of nanoparticles dispersed in about 5 ml ethanol on carbon-coated TEM grids.

**Zeta potential measurements**

Zeta potential measurements using a Zetasizer (Malvern ZS-90) were done after thorough sonication of the nanoparticles dispersed in different dispersants. Approximately 8 mg nanoparticles were dispersed in 15 ml of the dispersant, say ethanol, hexane and trioctylphosphine for a typical run. The number of runs made was in the range 50-100.

## Results and discussion

### Study of nanolattice formation through SAXS, XRD and TEM

The small angle X-ray scattering is a powerful tool to identify the nanolattice structures.[7] These data for the nickel nanoparticles of different particle sizes (~4-10nm) prepared in oleylamine and trioctylphosphine, without any other *extra* surfactant, reagents or external forces are shown in Figure 1. Several low angle peaks clearly observed in the SAXS data are assigned to the lattice planes formed by the monodispersed nanoparticles. Since no extra forces are used to prepare them, they indicate the natural formation of nanolattices that are distinct from wide angle X-ray diffraction (XRD) as the latter is due to the atomic face-centered cubic (fcc) lattice (Figure 2, left panel, and S1 in the Electronic Supplementary Information (ESI)). The selected area electron diffraction patterns of the nanoparticles for the electron beam perpendicular (Figure 1a, bottom left inset) and parallel to the plane of TEM grid plane (Figure 1a, bottom middle inset) for 10 ml sample reveal *local* self-assembly of hcp lattice of nanoparticles in two-dimensions. Figure 1a, bottom right inset shows an expanded TEM image of hexagonal arrangement of seven nanoparticles of nearly spherical shapes. Figure 2, righ panel, shows the typical TEM images of four samples of monodispersed nanoparticles. The statistical distribution plots of particle sizes (Figure 2, right panel, upper insets) indicate their monodispersed nature and respective average size. The selected area electron diffraction of the lattice (Figure 2, right panel, lower insets) confirms the fcc structure of the atomic lattice seen from XRD (Figure 2, left panel).

To ensure that the peaks in Figure 1 b, c and d are due to self-assembly of bulk 3D hcp structure of nanoparticles, *analytical calculations*[26] were made using these peaks. For this, $\sin^2\theta$ values were determined from

$$\sin^2\theta = A(h^2 + hk + k^2) + Cl^2, \qquad (1)$$

where $A=\lambda^2/3a^2$, $C=\lambda^2/4c^2$, $\lambda$ is wavelength of the X-ray and other parameters have their usual meanings. Permissible values of $(h^2+hk+k^2)$ being 1, 3, 4, 7, 9, etc for hcp structure, the observed $\sin^2\theta$ values were divided by 1, 3, 4, etc. These numbers were examined to find out that any of the quotients (nearly) match the observed $\sin^2\theta$ values and hence the tentative value of A was determined. The correspondingly matched values of (hk0) were chosen as the expected (hk0) values. Using these (hk0), and $A(h^2+hk+k^2)$ values, value of C is determined from equation (1) such that $Cl^2$ is in the ratio of 1, 4, 9, 16, etc. This procedure readily enables to identify the peaks in the pattern systematically. Final check was done by a comparison of observed and calculated $\sin^2\theta$ values.

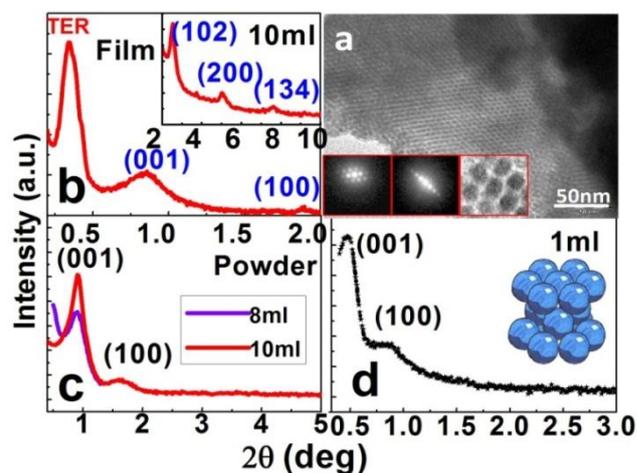

**Fig. 1** Transmission electron microscopy (TEM) images and SAXS patterns of Ni nanoparticles. (a) Representative TEM image of 10ml sample nanoparticles. Inset: SAED of hexagonally arranged self-assembled Ni nanolattice when the electron beam is perpendicular (left) and parallel (middle) to the plane of copper grid, and magnified portion of seven (hexagonally arranged) nanoparticles (right). (b) SAXS of glass-drop-casted film of 10ml sample with higher angle in inset. The analytically calculated (hkl) values for hcp phase are given; total external reflection (TER) is due to glass substrate. (c) Powder SAXS for 10ml and 8ml Ni bulk nanoparticle samples. (d) Powder SAXS of 1ml Ni bulk nanoparticle sample. Inset: an illustration of hexagonal closed-packed unit cell representing the nanoparticle unit cell.

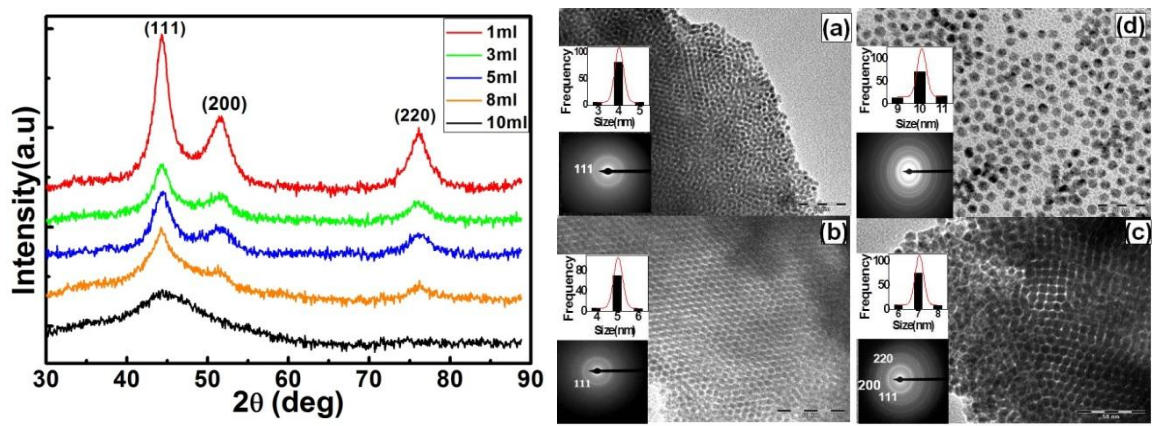

**Fig. 2** X-ray diffraction and transmission electron microscopy data. Left panel, XRD of nickel nanoparticles prepared for 1 ml, 3ml, 5ml, 8ml and 10ml trioctylphosphine content for fixed nickel acetylacetonate and oleylamine concentrations. Diffraction angle was converted to 1.5406 Å wavelength equivalent of copper to compare the laboratory XRD results. Right panel, TEM images of (a) 10ml, (b) 8ml, (c) 3ml and (d) 1ml sample nanoparticles. Insets: The statistical distribution plots of particle sizes and their fits (upper), and selected area electron diffraction of the atomic lattice (lower).

The SAXS peaks of the 10ml sample can thus be systematically correlated with (hkl) values of bulk 3D hcp nanolattice (Figure 1b) that has the nanolattice parameters, a=3.812nm, c=7.131nm and c/a=1.87. The peak near 2θ=0.36° was identified as the total external reflection from glass substrate (Figure S2 in the ESI). In contrast, our attempt to find out the peak positions for their probable fcc nanolattice using the average particle size of 4.0 nm as lattice spacing were always different from those observed. This proves that the observed SAXS peaks are due to bulk 3D hcp nanolattice, not due to fcc nanolattice. The peaks of the other samples were also identified as hcp nanolattice. The 10 ml and 8 ml samples in powder forms show hexagonal structure (Figure 1c) with a, c and c/a of 4.39 nm & 4.4 nm, 7.31 nm & 7.45 nm and 1.67 & 1.69, respectively. For 10 ml powder sample, the nanolattice parameters are slightly bigger than those of thin film. The nanolattice parameters of 1 ml powder sample (Figure 1d) are a=8.09 nm, c=13.28 nm with c/a=1.64. These natural bulk hcp nanolattices even in powder form are striking. They imply that such nanolattices should prevail even in compacted pellets as well as in refs. [27,28], similar to sample powder of atomic lattices.[17,18,26] The ratio c/a=1.64-1.87 found is similar to atomic lattices indicating their close analogy. Notably, value of the nanolattice parameter a is smaller in some of nanolattices than the average particle size. This is explained on the basis of adjustable cappant thickness.[8,9]

In order to comprehensively establish the genuine origin of formation of the natural nanolattice is due to the use of trioctylphosphine or any other surfactant, we have recorded the SAXS patterns (Figure S3 in the ESI) of the several other samples prepared for (i) varying oleylamine with fixed trioctylphosphine, (ii) varying triphenylphosphine with fixed oleylamine and (iii) separately for each of these surfactants. Remarkably, the SAXS peaks associated with the nanolattice formation is naturally observed whenever trioctylphosphine only or in combination of it with other surfactants are used for the sample preparations, but not with triphenylphsophine and oleylamine separately. The nanolattice observed is therefore ascribed to the trioctylphosphine that in turn also is expected to prove the vestiges of nanolattices seen in ZnS and CdSe[24] and nickel[6] nanoparticles as due to the trioctylphosphine. The natural cohesive energy of the nanolattice is attributed to the dissociated molecules or impurities of trioctylphosphine.[15,16] The large clusters of these nanolattices for average particle sizes of 4nm, 5.1nm, 7.1nm and 10.1nm can be clearly seen from TEM images in bigger scales (Figure S4 in ESI).

### Zeta potential properties

To ascertain the stability of these nanolattices and prove its formability, we have carried out their zeta potential (ζ) measurements. We note that ethanol might assume crucial if the formation of nanolattice was also related to it as the nanoparticles were washed with or dispersed in it. The ζ data for samples prepared in trioctylphosphine and oleylamine together measured in ethanol, hexane and trioctylphosphine separately are represented in Figure 3A. These ζ values in the range of -1 to 1.5mV in ethanol are quite small while those of -19 to -38 mV in trioctylphosphine are relatively large. This indicates that influence of ethanol to trioctylphosphine ligands on the electric double layers of the particles is marginal while that of trioctylphosphine dispersant to its bound counterparts on the surface of nanoparticles is significantly large that increases with its number of ligands. The trioctylephosphine concentration normally leads to reduction in particle size (Figure 3A, Size curve). The situation for hexane is however very random (Table 1). The small values of ζ indicate the weak ionic nature of surfactants that in turn seem to confirm the formation of aggregates i.e. nanolattice.[16] The weak ionic nature combined with high values of conductivity and mobility of the nanoparticles in ethanol, 0.392 - 10.2 µS/cm and -0.1768 - 0.1148 µmcm/Vs, respectively (Table 1) would indicate that these non-ionic surfactants are quite ionic. They are attributed to cation-anion pairs formed from the minute contaminants or dissociation of trioctylphosphine molecules.[15,16] This leads to the formation of nanolattice, well in agreement with the Bjerrum radius[16] (~28nm) that is much larger than the size of present nanoparticles (4-10nm).

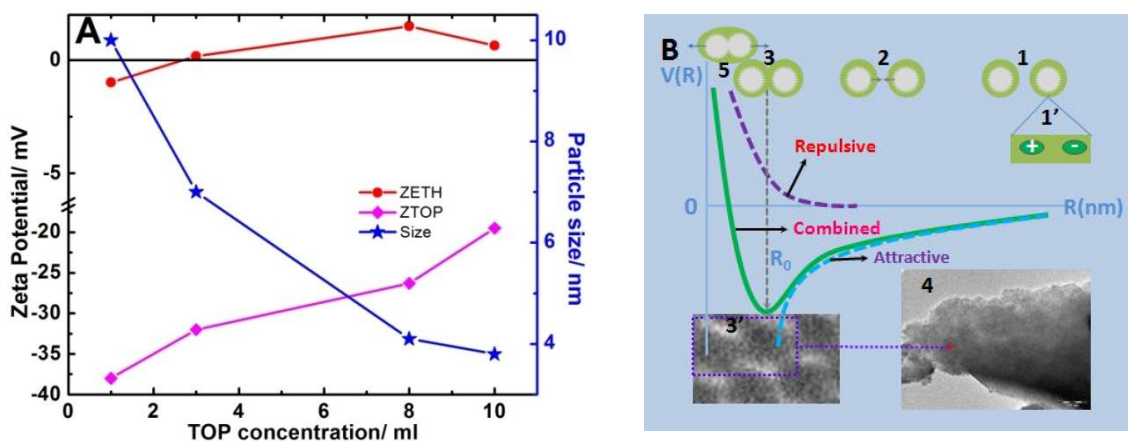

**Fig. 3** Zeta potential of various nanoparticle sizes in different dispersants and the proposed pair potential. A, Zeta potential in ethanol (ZETH) and trioctylphosphine (ZTOP) for various trioctylphosphine concentrations. Various dispersants were used to see their influence on zeta potential considering the media within which samples were prepared or treated later after preparation as that might influence the nanolattice. Right axis shows the particle size variation with trioctylphosphine concentrations that show increase in particle size (Size curve) as the concentration (in ml) of the latter decreases. B, Schematic plot of pair potential. The expected Coulombic nature is shown for $R > R_0$ (1,2) until it reaches an equilibrium position $R_0$ (3) and infinitely repulsive for $R < R_0$ (5). Each particle is surrounded by (green shell) either several surfactant molecules or ion pairs with negative and positive charges (1'). At the equilibrium position $R_0$ (3 or 3') nanolattice (4) is formed; scale in inset (4) is 100nm.

### Mechanism of the nanolattice formation

Clearly, natural nanolattice is formed when non-ionic long- and triple-chained trioctylphosphine is used as surfactant, not with that of long-chain (oleylamine) or phenyl group (triphenylphosphine) surfactant. This nanolattice formation is tentatively understood in two ways. First, a head of surfactant (P or trioctylphosphine) binds surface of Ni particle while organic tail in turn binds tail of another surfactant so that particles are glued at a fixed distance. When the number of such processes increases, nanolattice formation takes place with the minimization of the total surface energy. Secondly, according to zeta potential data, cations of the ion pairs (Figure 3B, inset 1') of the dissociated surfactant molecules or impurities attached on a nanoparticle will attract the anions of the surrounding nanoparticles until they sense the presence of other cations of the latter leading to a repulsion. This in analogy with the electrons and nuclei of atoms in an atomic lattice[17,18] leads to an attractive pair potential combined with repulsive potential. This is illustrated by nanoparticle pairs with their separation R (Figure 3B). The resultant potential binds the nanoparticles enabling the observed equilibrated nanolattice (Figure 3B, inset 4). The pair potential will therefore be Coulombic for $R > R_0$ until it reaches an equilibrium position $R_0$ (Figure 3B, inset 3 or 3') and infinitely repulsive for $R < R_0$ (Figure 3B, inset 5). The present results, establishing clearly the natural formation of bulk hcp nanolattices when trioctylphosphine is used as surfactant and consequent resultant cohesive energy, are therefore striking. They are distinct from earlier reports that use an external energy or extra media[6-10], and show the unique property of trioctylphosphine as a creator of nanolattice. The cohesive energy considered here is expected to include all other cohesive energies that may arise.[17,18] The example of trioctylphosphine as a former of nanolattice of at least Ni[6], ZnS and CdSe[24] clearly shows that this surfactant may be used to grow varieties of natural nanolattices of choice and the similar approach may be applied to other surfactants to enable natural nanolattice formation. Therefore, the enhanced luminescence in PbS[23] and in ZnS/ZnS and CdSe[24] is likely to be related to their nanolattices being formed.

### Conclusions

In conclusion, we have successfully prepared monodispersed nickel nanoparticles of different sizes in the range of ~4-10nm taking trioctylphosphine as one of the surfactants. Such nanoparticles form outstandingly natural hexagonal close-packed nanolattices without external forces for the nanoparticles prepared as usual whenever trioctylphosphine is (one of) the surfactant(s). The nanolattice parameters, calculated analytically, have c/a ratios identical to those of atomic lattices. Moreover, these results undoubtedly establish the exceptional role of nanolattice formability of trioctylphosphine of several materials including, but not limited to, nickel, ZnS and CdSe. The nanolattice formability is explained based on the balanced attractive and repulsive energies of cation-anion pairs of the dissociated surfactant molecules or impurities. These findings will therefore provide a far-reaching new outlook for research in desired natural nanolattices for other similar surfactants as well, without using extra forces, and for understanding their properties for varieties of future applications.

## Acknowledgements

Authors thank the Department of Science and Technology, India for the financial support and Saha Institute of Nuclear Physics, India for facilitating the experiments (especially Sanjay Singh) at the Indian Beamline, Photon Factory, KEK, Japan; M. Gupta and V. R. Reddy, UGC-DAE CSR, Indore for laboratory XRD and HR-SAXS data, respectively; and S. Satapathy, Raja Ramanna Centre for Advanced Technology, Indore, India for zeta potential data. They also acknowledge thankfully discussions with Rajendra Prasad, Devi Ahilya University, Indore; A. Sundaresan, Jawaharlal Nehru Centre for Advanced Scientific Research, Bangalore; Raghumani S. Ningthoujam, Bhabha Atomic Research Centre, Mumbai, India; and Xavier Crispin, Linkopin University, Sweden.

## Notes and references

[a] Department of Physics, ISLE, IPS Academy, Rajendra Nagar, Indore (MP) 452012, India.

[b] Department of Physics, Government Holkar Science College, A. B. Road, Indore (MP) 452001, India.

[c] Crystallography Laboratory, UGC-DAE Consortium for Scientific Research, Khandwa Road, Indore (MP) 452001, India.

[d] Electrical Transport Laboratory, UGC-DAE Consortium for Scientific Research, Khandwa Road, Indore (MP) 452001, India.

*Email: okram@csr.res.in, okramgs@gmail.com; Tel: +91-731-2463913, 2762267

Electronic Supplementary Information (ESI) available: [(extra XRD to compare laboratory and synchrotron results, laboratory high resolution small angle X-ray scattering data and transmission electron microscopy images]. See DOI: 10.1039/c000000x/

1 J. M. Luther, M. Law, Q. Song, C. L. Perkins, M. C. Beard and A. J. Nozik, *ACS Nano*, 2008, **2**, 271.
2 F. X. Redl, K. S. Cho, C. B. Murray and S. O'Brien, *Nature*, 2003, **423**, 968.
3 S. Sun, C. B. Murray, D. Weller, L. Folks and A. Moser, *Science*, 2000, **287**, 1989.
4 Z. Nie, A. Petukhova and E. Kumacheva, *Nature Nanotech*, 2010, **5**, 15.
5 M. P. Pileni, *J. Phys. Chem.*, 2001, **B105**, 3358.
6 J. Park, E. Kang, S. U. Son, H. M. Park, M. K. Lee, J. ; Kim, K. W. Kim, H. J. Noh, J. H. Park, C. J. Bae, J. G. Park and T. H. Park, *J. Adv. Mater*, 2005, **17**, 429.
7 D. Nykypanchuk, M. M. Maye, D. V. Lelie and O. Gang, *Nature,* 2008, **451**, 549.
8 Y. Min, M. Akbulut, K. Kristiansen, Y. Golan and J. Israelachvili, *Nature Mater*, 2008, **7**, 527.
9 K. J. M. Bishop, C. F. Wimer, S. Soh and B. A. Grzybowski, *Small*, 2009, **5**, 1600.
10 Y. Gao, Y. Bao, M. Beerman, A. Yasuhara, D. Shindo and K. Krishnan, *Appl. Phys. Lett.*, 2004, **84**, 3361.
11 E. V. Shevchenko, D. V. Talapin, N. A. Kotov, S. O'Brien and C. B. Murray, *Nature* 2006, **439**, 55.
12 X. M. Lin, H. M. Jaeger, C. M. Sorensen and K. J. Klabunde, *J. Phys. Chem.*, 2001, **105**, 3353.
13 T. Hyeon, S. S. Lee, J. Park, Y. Chung and H. B. Na, *J. Chem. Soc.*, 2001, **123**, 12798.
14 M. E. Leunissen, C. G. Christova, A. P. Hynninen, C. P. Royall, A. I. Campbell, A. Imhof, M. Dijkstra, R. V. Roij and A. V. Blaaderen, *Nature*, 2005, **437**, 235.
15 A. S. Dukhin and P. J. Goetz, surfactants in non-polar liquids, Dispersion Technology, Inc. New York 2014. http://www.dispersion.com/ionic-properti es-of-so-called-non-ionic-surfactants-in-non-polar-liquids (accessed May 20, 2014).
16 N. Bjerrum, Proceedings of the 7th International Congress of Applied Chemistry, London 1909, *Section X*, pp .55-60.
17 C. Kittel, *Solid State Physics*, 7th ed., John Wiley and Sons Inc: New York, 1996, p 55-95.
18 N. W. Ashcroft and N. D. Mermin, *Solid State Physics*, Saunders College Publishing, New York, 1976, p 395-414.
19 S. Carenco, C. Boissiere, L. Nicole, C. Sanchez, P. L. Floch andN. Mezailles, *Chem. Mater.*, 2010, **22**, 1340.
20 S. Chen, X. Zhang, Q. Zhang and W. Tan, *Nanoscale Res. Lett.*, 2009, **4**, 1159.
21 S. Sharma, N. S. Gajbhiye and R. S. Ningthoujam, *J. Colloid Interface Science*, 2010, **351**, 323.
22 N. S. Gajbhiye, S. Sharma, A. K. Nigam and R. S. Ningthoujam, *Chem. Phys. Lett.*, 2008, **466**, 181.
23 K. A. Abel, J. Shan, J. C. Boyer, F. Harris and Frank C. J. M. van Veggel, *Chem. Mater.*, 2008, **20**, 3794.
24 D. V. Talapin, A. L. Rogach, A. Kornowski, M. Haase and H. Weller, *Nano Lett.*, 2001, **4**, 207.
25 T. Mokari, M. Zhang and P. Yang, *J. Am. Chem. Soc.*, 2007, **129***,* 9864.
26 B. D. Culity, *Elements of X-Ray Diffraction*, Addison-Wesley, Massachusetts, 1956.
27 A. Soni and G. S. Okram, *Appl. Phys. Lett.*, 2009, **95**, 013101.
28 G. S. Okram and Netram Kaurav, *J. Appl. Phys.*, 2011, **110**, 023713.

**Table 1** Zeta potential (ζ) data with conductivity (σ) and mobility parameters ($\mu_H$) in ethanol, hexane and trioctylphosphine

| Sample (In TOP concentration) | Ethanol | | | Hexane | | | Trioctylphosphine | | |
|---|---|---|---|---|---|---|---|---|---|
| | ζ (mV) | σ (mS/cm) | $\mu_H$ (μmcm/Vs) | ζ (mV) | σ (mS/cm) | $\mu_H$ (μmcm/Vs) | ζ (mV) | σ (mS/cm) | $\mu_H$ (μmcm/Vs) |
| 1ml | -0.986 | 0.000392 | -0.1768 | 151/125 | --- | 0.8448 | -38.0 | 2.78e-4 | -0.01584 |
| 3ml | 0.173 | 0.00323 | 0.03098 | -5.9/25 | 4.03e-4 | -0.03307 | -29.9 | 1.48e-4 | -0.01245 |
| 8ml | 3.29 | 0.0102 | 0.5908 | -102/54.7 | 3.82e-4 | -0.5718 | -26.3 | 1.81e-4 | -0.01097 |
| 10ml | 0.639 | 0.00731 | 0.1148 | 6.51 | 6.73e-4 | 0.03649 | -19.5 | 3.41e-4 | -0.008115 |
| Viscosity | 0.12cP | | | 0.2970cP | | | 10.2cP | | |

# Electronic supporting material

# Naturally self-assembled nickel nanolattice

Jaiveer Singh,[a] Netram Kaurav,[b] Niranjan Prasad Lalla[c] and Gunadhor Singh Okram[*d]

[a]Department of Physics, ISLE, IPS Academy, Rajendra Nagar, Indore (MP) 452012, India.
[b]Department of Physics, Government Holkar Science College, A. B. Road, Indore (MP) 452001, India.
[c]Crystallography Laboratory, UGC-DAE Consortium for Scientific Research, Khandwa Road, Indore (MP) 452001, India.
[d]Electrical Transport Laboratory, UGC-DAE Consortium for Scientific Research, Khandwa Road, Indore (MP) 452001, India.
*Email: okram@csr.res.in, okramgs@gmail.com; Tel: +91-731-2463913,2762267

This electronic supporting information section consists of extra XRD to compare laboratory and synchrotron results, laboratory high resolution small angle X-ray scattering data, and transmission electron microscopy images.

## 1. Characterization
### 1.1 X-Ray diffraction

Samples were characterized using both the laboratory and synchrotron radiation X-ray diffraction (XRD) as in Figure S1. It is clearly seen that three peaks due to (111), (200) and (220) planes are visible in Laboratory XRD for 1ml and 3ml samples only, whereas they are seen in synchrotron XRD for all samples except 10ml sample. These results combined with SAED demonstrate clearly that the synchrotron data is superior to those of laboratory XRD and electron diffraction viz. SAED (Fig.2).

### 1.2 High resolution X-ray diffraction (Laboratory)

High resolution laboratory XRD was done using Cu K$\alpha$ radiation in the angle range 0.2-10° with a step size of 0.02°. This was basically done for ascertaining preliminary and final information that "Whether we can observe the nanolattice of nickel nanoparticles when drop-casted on a glass slide as seen in SAED or not (Figure 1a)." It is evident that such possibility prevails as indicated by the prominent four peaks (Figure S2). There is strong proof that nickel nanoparticles form a bulk 3D nanolattice as discussed in the text. The first peak near $2\theta=0.36°$ was identified as the total external reflection from glass substrate (Figure S2). This was confirmed from similar experiment on blank glass slide that exhibits only one peak due to total external reflection.

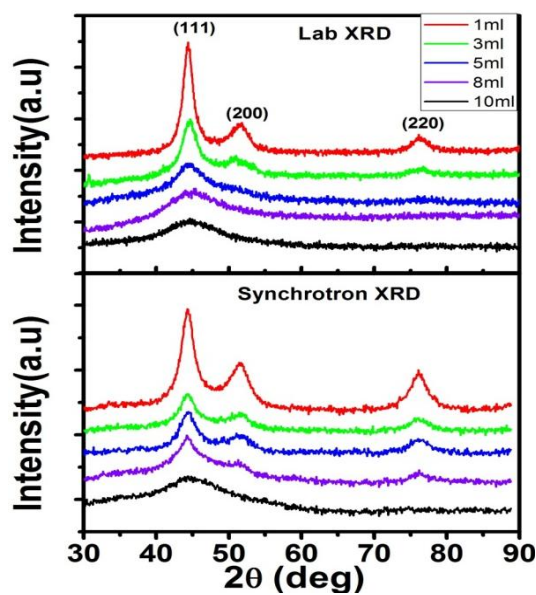

**Figure S1** A comparison of the X-ray diffraction data of the samples indicated for the laboratory and synchrotron radiation sources. Synchrotron data clearly reveals the peaks at (200) and (220) even for 8ml sample, which are otherwise not seen in laboratory XRD down to 3ml sample.

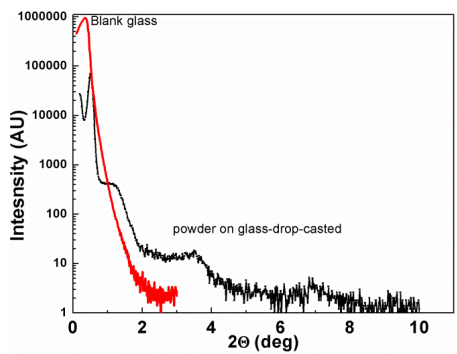

**Figure S2** Grazing incident X-ray diffraction of (Cu K$_\alpha$) laboratory source for 10ml sample drop-casted film on glass slide (black) and blank glass slide (red) as indicated. This confirms that the first peak near 0.36° is due to the total external reflection. Note: Angle of incidence was fixed at 0.25°.

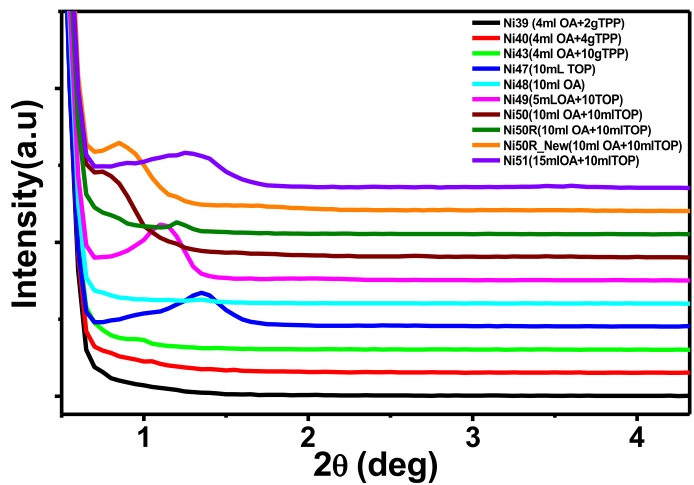

**Figure S3** Small angle X-ray scattering data of various samples. Samples with trioctylphosphine (combined with oleylamine also) surely exhibit peaks, indicating the formation of nanolattice while those with Olyelamine (OA), triphenylphosphine (TPP) or combined do not exhibit any peak evidencing that these surfactants do not support formation of nanolattice. This shows dominant role of trioctylphosphine in enabling the nanolattice formation; only one peak is seen in these SAXS patterns as the set up setting the incident X-ray angle had to be done below 0.5°, not like that in Figure S2. Had we chose smaller angle of incidence, more peaks would have been expected.

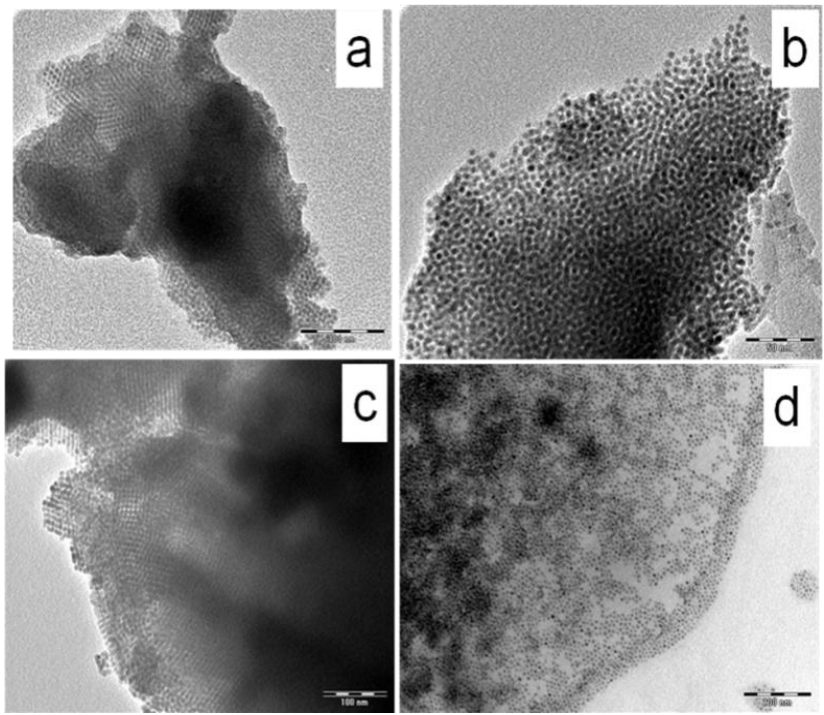

**Figure S4** Representative Transmission Electron Microscopy images showing natural self-assembly. Natural self-assembly of (a) 10ml, (b) 8ml, (c) 3ml and (d) 1ml trioctylphosphine prepared nanoparticle samples for fixed 8ml oleylamine. Scales shown are for 100nm, 50nm, 100nm and 200nm, respectively.